\documentclass{PoS}

\usepackage{amsmath}
\newcommand{\cutoff}{p_{T, \textrm{\footnotesize cutoff}}^\gamma}

\title{Centrality dependence of the direct photon multiplicity in heavy ion collisions}

\ShortTitle{Centrality dependence of the direct photon multiplicity in heavy ion collisions}

\author{Charles Gale\\
        Department of Physics, McGill University, 3600 University Street, Montr\'eal, QC, H3A 2T8, Canada\\
        E-mail: \email{gale@physics.mcgill.ca}}
    
\author{Sangyong Jeon\\
	Department of Physics, McGill University, 3600 University Street, Montr\'eal, QC, H3A 2T8, Canada\\
	E-mail: \email{jeon@physics.mcgill.ca}}

\author{Scott McDonald\\
	Department of Physics, McGill University, 3600 University Street, Montr\'eal, QC, H3A 2T8, Canada\\
	E-mail: \email{scott.mcdonald2@mail.mcgill.ca}}

\author{\speaker{Jean-Fran\c cois Paquet}\\
	Department of Physics, Duke University, Durham, NC 27708, USA\\
	E-mail: \email{jeanfrancois.paquet@duke.edu}}

\author{Chun Shen\\
Department of Physics and Astronomy, Wayne State University, Detroit, Michigan, USA \\
RIKEN BNL Research Center, Brookhaven National Laboratory, Upton, NY 11973, USA\\
	E-mail: \email{chunshen@wayne.edu}}

\abstract{Measurements indicate that the centrality dependence of the direct photon multiplicity scales approximately with the number of binary nucleon collisions. Importantly, these same measurements suggest that this scaling does not depend on the cutoff used to define the photon multiplicity. In this work we provide a theoretical perspective on these observations. 
We emphasize the importance of stronger experimental constraints, in particular for the cutoff independence of the centrality scaling. We highlight the importance of revisiting theoretical studies of jet-medium photons and of medium-modified prompt photons to clarify their contribution to the direct photon multiplicity, as well as the magnitude and direction of their momentum anisotropy.}

\FullConference{International Conference on Hard and Electromagnetic Probes of High-Energy Nuclear Collisions\\
		30 September - 5 October 2018\\
		Aix-Les-Bains, Savoie, France}

\begin{document}



An important source of photons in nuclear collisions are ``prompt photons''. They are produced in initial hard parton collisions, through two mechanisms\footnote{This division into two mechanisms is useful conceptually, although it can only be done rigorously at leading order.}
: 
	(i) processes such as Compton scattering ($q g \to q \gamma$) 
	and quark annihilation ($q \bar{q} \to g \gamma$), 
	where a photon is a direct final state of the hard parton interaction; and
	(ii) fragmentation, where a hard parton-parton collision (e.g. $q \bar{q} \to q \bar{q}$) is followed by a photon being radiated from one of the final state partons.

Prompt photon production in proton-proton (p+p) collisions has been calculated in perturbation theory at next-to-leading order in $\alpha_s$;
agreement with p+p data is good
in a wide range of center-of-mass energies~\cite{Aurenche:2006vj}. 
Prompt photons remain an important source of direct photons in heavy ion collisions. They are generally said
to scale with the number of binary nucleon collisions, up to modest corrections from isospin and nuclear effects on the parton distribution functions; this statement is supported by the good agreement of high $p_T^\gamma$ photon spectra ($\gtrsim 5-10$~GeV)  in heavy ion measurements with binary-scaled prompt photon calculations (or p+p photon data).
At low  $p_T^\gamma$, the production of prompt photon in heavy ion collisions is more complex.
The fragmentation component
of prompt photons, subdominant at high $p_T^\gamma$, is the dominant source of prompt photons at low $p_T^\gamma$. These fragmentation photons are affected by parton energy loss. 
Moreover prompt photons are not the only source of low $p_T^\gamma$ direct photons in heavy ion collisions: additional photons originate from (i) blackbody radiation produced by the hot expanding plasma (``thermal photons''), and (ii) ``jet-medium photons'' produced in parton-plasma interactions (interactions which also lead to parton energy loss) \cite{Turbide:2007mi}. 
Measurements indicate that the low $p_T^\gamma$ direct photon spectra measured in heavy ion collisions is considerably larger than binary-scaled photon spectra measurements from proton-proton collisions. This suggests that thermal and jet-medium photons more than compensate for the suppression of prompt photons resulting from parton energy loss. Additional simulations are still required to confirm this scenario\footnote{Almost all recent comparisons with data rely on prompt photon calculations that neglect the effect of energy loss.
Up-to-date calculations of jet-medium photons and of prompt photons that account for parton energy loss will be essential to clarify the status of model-to-data comparisons.
}.

Recent work by the PHENIX Collaboration brought the role of prompt photons back to the front of the discussion~\cite{Adare:2018wgc}. They observed that the \emph{centrality dependence} of the direct photon multiplicity is consistent with binary collisions scaling:
\begin{equation}
\left. d N_{direct}/d y \right|_{p_T^\gamma>\cutoff}
\approx N_{binary}^{\alpha} K\left(\sqrt{s_{NN}},\cutoff \right) 
\label{eq:phenix_scaling}
\end{equation}
where the number of binary collisions $N_{binary}$ is a 
calculated
from the Monte-Carlo Glauber model and the exponent $\alpha$ was found to be consistent with $1$. The normalization factor $K(\sqrt{s_{NN}},\cutoff )$ depends on the center-of-mass energy $\sqrt{s_{NN}}$, and the multiplicity cutoff $p_T^\gamma$. However the centrality dependence itself of the multiplicity appears not to depend on the cutoff $\cutoff$ (i.e. $\alpha\approx 1$ independent of $\cutoff$).
The same conclusions had already been obtained for one center-of-mass energy in Ref.~\cite{Adare:2014fwh}. 

In what follows we provide a theoretical perspective on this binary-scaling from the point of view of thermal photons, medium-modified prompt photons and jet-medium photons.


\paragraph{Medium-modified prompt photons and jet-medium photons}
The multiplicity of sufficiently high $p_T^\gamma$ prompt and jet-medium photons in heavy ion collisions can be written schematically as
\begin{align}
\left. d N_{j}/d y \right|_{p_T^\gamma>\cutoff} =
\frac{N_{binary}}{\sigma^{inel}_{NN }} & \int \frac{d \phi}{2 \pi} \int_{\cutoff} d p_T p_T \left[ \sum_{a, b, c} f_{a/A} \left(x_a,Q\right) 
\otimes  f_{b/A}\left(x_b,Q \right) \otimes \, d\hat{\sigma}_{a b \to c \gamma}  \right. \nonumber \\
& \left. +\sum_{a, b, c, d} f_{a/A} \left(x_a,Q\right) 
\otimes  f_{b/A}\left(x_b,Q \right) \otimes \, d\hat{\sigma}_{a b \to c d}  \otimes D^M_{\gamma/c} \left(z_c,Q\right)
\right]
\label{eq:prompt_jetmedium}
\end{align}
where $f(x,Q)$ are nuclear parton distribution;
$d\hat{\sigma}$ are perturbative parton cross-sections; and $\otimes$ represents a convolution over the kinematic variables. All effects from the quark-gluon plasma --- parton energy loss and jet-medium photon production --- are absorbed into a medium-modified fragmentation function $D^M_{\gamma/c} \left(z_c,Q\right)$. At a fixed collision energy, the centrality dependence of  Eq.~\ref{eq:prompt_jetmedium} originates from the $N_{binary}$ pre-factor as well as from $D^M_{\gamma/c} \left(z_c,Q\right)$. This medium-modified fragmentation function is assumed to encode the same non-perturbative fragmentation into photons as the vacuum function. However, the perturbative sector of the vacuum fragmentation function is modified to account for the presence of a medium: it includes\footnote{We are not aware of numerical studies of photons that include both medium-modified high and low-virtuality showering. Simulations that include only the latter have been performed previously, however (e.g. Ref.~\cite{Turbide:2005fk}).} (i) perturbative DGLAP-like photon emission between virtuality $Q$ and a lower virtuality $Q_0$; and (ii) perturbative low-virtuality showering, which simultaneously leads to medium-induced photon emission and medium-induced parton energy loss. In vacuum, only the  DGLAP-like photon emission is present.
The final non-perturbative fragmentation into photons is performed after this perturbative showering. Unlike vacuum fragmentation functions, the medium-modified $D^M_{\gamma/c} \left(z_c,Q\right)$ is not universal: it must be evaluated dynamically to account for the exact profile of the medium.


Typically the spectra of prompt and jet-medium photons fall off rapidly with $p_T^\gamma$ and the photon multiplicity is dominated by the smallest $p_T^\gamma$ regions of the integrand ($p_T^\gamma \sim \cutoff$).
This presents a challenge for Eq.~\ref{eq:prompt_jetmedium}. 
Intrinsically in collinear perturbative calculations, there is a lower scale $Q_0 \sim 1-2$~GeV separating the perturbative and non-perturbative sectors. In Eq.~\ref{eq:prompt_jetmedium}, the scale $Q>Q_0$ is expected to be of the order of the photon transverse momentum: $Q\sim p_T^\gamma$. This implies that multiplicities evaluated with small $\cutoff\sim Q_0$ may be dominated by $p_T^\gamma$ regions where the perturbative calculations are the least reliable. Careful numerical simulations of medium-modified prompt and jet-medium photons will be essential to clarify the situation. Alternative approaches to calculate low momentum jet and photon production may also be necessary (see e.g. Ref.~\cite{Hattori:2016jix} and references therein).

\paragraph{Thermal photons}
Thermal photons are calculated by convoluting a thermal emission rate $E d^3 \Gamma/d^3 p$ with a spacetime profile of the plasma obtained from hydrodynamics:
\begin{equation}
\left. d N_{th}/d y \right|_{p_T^\gamma>\cutoff} =
\int d^4X \int \frac{d \phi}{2 \pi} \int_{\cutoff} d p_T p_T  \left [E \frac{d^3 \Gamma}{d^3 p}\left(P \cdot u(X),T(X),\pi^{\mu\nu}(X),\Pi(X)\right) \right]
\label{eq:thermal}
\end{equation}
where $T$, $u$, $\pi^{\mu\nu}(X)$,$\Pi(X)$ are the temperature, flow velocity, shear tensor and bulk pressure profiles of the plasma, and $P$ is the photon four-momentum.

\begin{figure}[t]
	\includegraphics[width=0.38\linewidth]{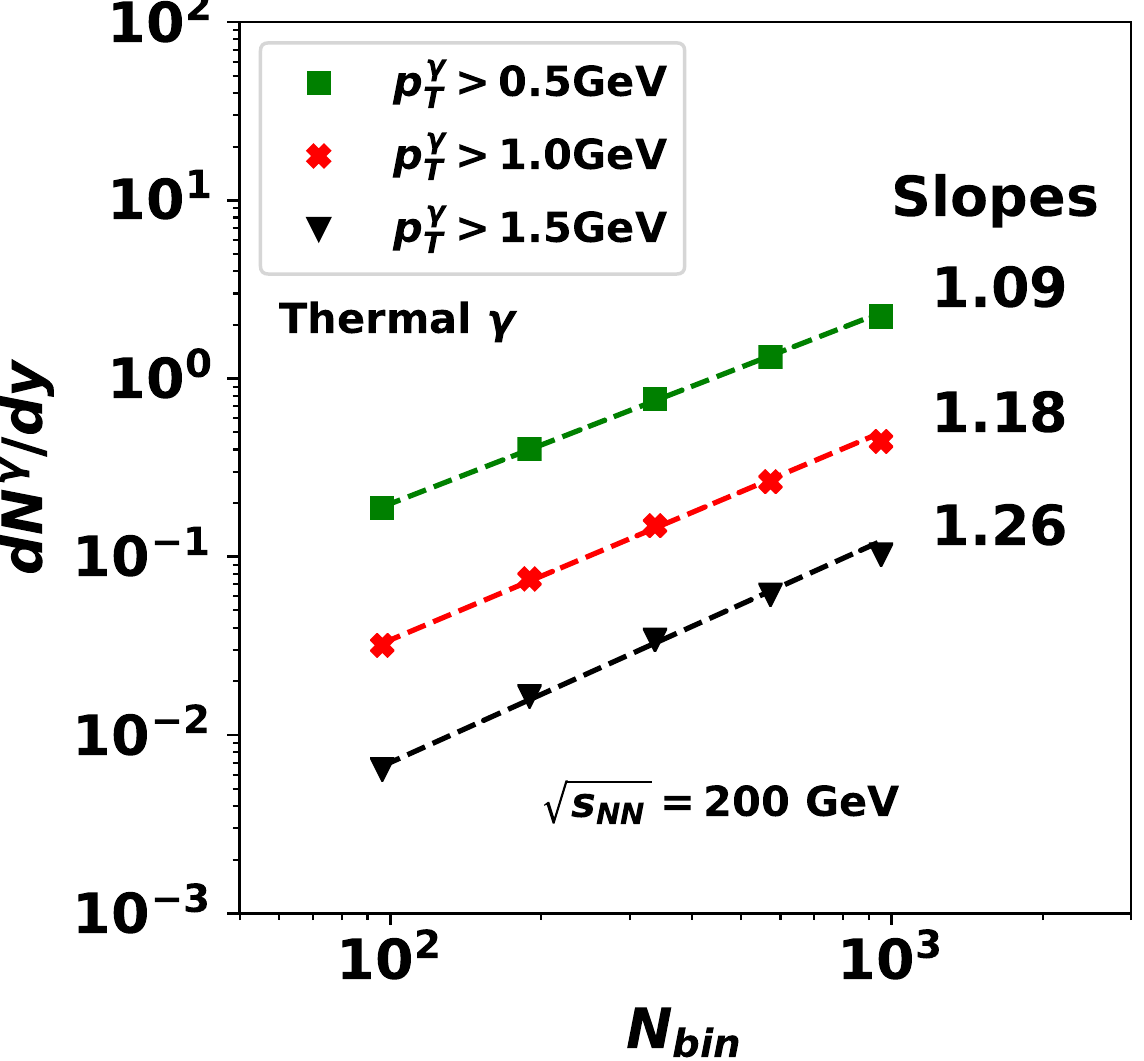}
	\hspace{2.cm}
	\includegraphics[width=0.39\linewidth]{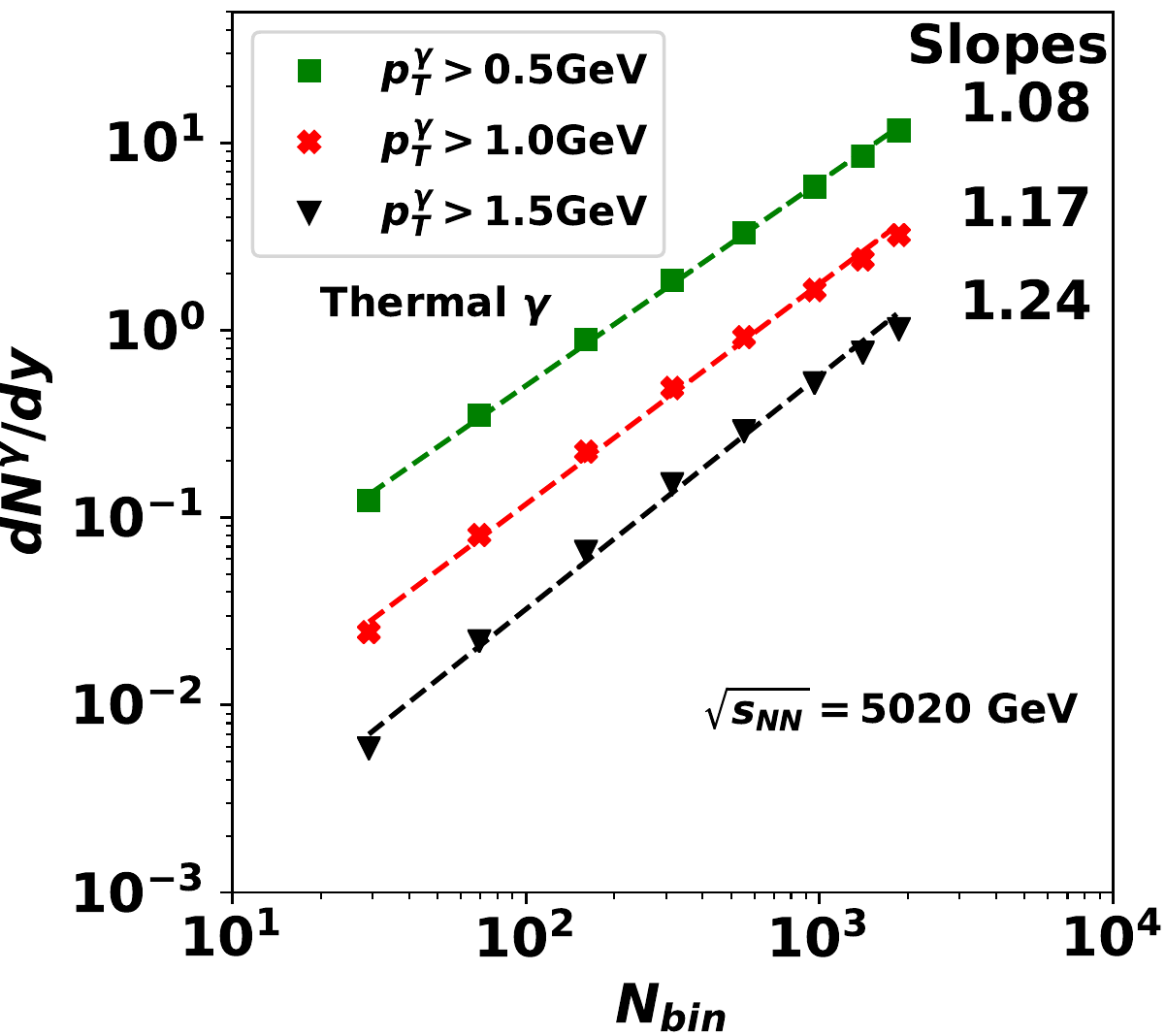}
	\caption{Thermal photon multiplicity as a function of the number of binary nucleon collisions, obtained by varying the centrality for (a) Au-Au at $\sqrt{s_{NN}}=200$~GeV, and (b) Pb-Pb at $\sqrt{s_{NN}}=5020$~GeV.}
	\label{fig:thermal_vs_bin}
\end{figure}

Figure~\ref{fig:thermal_vs_bin} shows the power law scaling of the photon multiplicity as a function of $N_{binary}$
, for Au-Au collisions at $\sqrt{s_{NN}}=200$~GeV and Pb-Pb collisions at $\sqrt{s_{NN}}=5020$~GeV. Three different multiplicity cutoff  $\cutoff$ are shown: 0.5, 1 and 1.5~GeV. The results can be fitted reasonably well with a linear function; the slopes
are indicated on the figures. Based on Figure~\ref{fig:thermal_vs_bin}, we can write the thermal photon multiplicity as:
\begin{equation}
\left. d N_{th}/d y \right|_{p_T^\gamma>\cutoff}
\approx N_{binary}^{\alpha\left(\sqrt{s_{NN}},\cutoff\right)} M\left(\sqrt{s_{NN}},\cutoff \right) 
\label{eq:thermal_Nbin_scaling}
\end{equation}
where $M(\sqrt{s_{NN}},\cutoff ) $ is a function that can be tabulated, and $\alpha(\sqrt{s_{NN}},\cutoff)$ is an exponent for $N_{binary}$ that is approximately $1.1-1.3$. This exponent increases as the cutoff $\cutoff$ is increased. The exact numerical values of $\alpha(\sqrt{s_{NN}},\cutoff)$ depend on the details of the hydrodynamic simulation (initial conditions, viscosities, \ldots) as well as the photon emission rates; this exact dependence will need to be studied in greater details\footnote{In particular, we believe an observation made in Ref.~\cite{Shen:2013vja} --- that an increase in the photon emission rate at low temperatures leads to smaller slopes for the centrality dependence --- may not as general as previously thought.}.
Nevertheless, the $\cutoff$ \emph{dependence} of these values is a robust prediction: there is no expected scenario where the thermal photon multiplicity is independent from this cutoff.

\paragraph{Discussion \& Outlook}


Any discussion of the multiplicity scaling must be made with other photon observables in mind, in particular the photon momentum anisotropy $v_2$. Thermal photons generally have a large $v_2$ at low $p_T^\gamma$, while medium-modified prompt and jet-medium are generally understood to have a small $v_2$~\cite{Turbide:2005bz}. The measured photon $v_2$ is large, and is interpreted as favoring thermal photons as the dominant source of low $p_T^\gamma$ direct photons.


Summing thermal, prompt and jet-medium photons, we write the direct photon multiplicity:
\begin{equation}
\left. d N_{direct}/d y \right|_{p_T^\gamma>\cutoff}
\approx  \left[ 
\left. d N_{j}/d y \right|_{p_T^\gamma>\cutoff}
+ N_{binary}^{\alpha\left(\sqrt{s_{NN}},\cutoff\right)} M\left(\sqrt{s_{NN}},\cutoff \right)  \right]
\end{equation}

Inevitably the centrality dependence of this prediction depends on the relative size of thermal photons as opposed to that of the medium-modified prompt photons and jet-medium photons. Given we do not currently have calculations of the latter two sources in a setting consistent with our thermal photon calculations, we limit our discussion to asymptotic scenarios.

The first limit is a multiplicity dominated by jet-medium and medium-modified prompt photons. To simultaneously describe the measured photon $v_2$, these jet-medium and prompt photons need to have a large $v_2$. This scenario is arguably not supported by older calculations~\cite{Turbide:2005bz}. Whether calculations that include recent developments in hydrodynamics simulations (e.g. fluctuating initial conditions, initial flow and bulk viscosity) would produce a significantly larger momentum anisotropy for these photons will need to be studied numerically.

If thermal photons dominate the multiplicity, the centrality dependence of the photon multiplicity should be close to $N_{binary}^{\alpha}$ with $\alpha=\alpha(\sqrt{s_{NN}},\cutoff)$ discussed above. Importantly measurements could rule out this possibility with improved constraints on the cutoff \emph{independence} of the $\alpha$ exponent.

Given that intermediate scenarios are also possible,
additional measurements and calculations are essential to determine if the observed centrality scaling can be explained by our current understanding of photon production, or if there is a new direct photon puzzle.

\paragraph{Acknowledgements} We are grateful to the ALICE and PHENIX photon working groups for insightful discussions. This work was supported by the U.S. Department of Energy
under Award Numbers DE-FG02-05ER41367 (JFP) and DE-SC0013460 (CS), and by the Natural Sciences and Engineering Research Council of Canada. SM acknowledges funding from The Fonds de recherche du Qu\'ebec - Nature et technologies 
(FRQ-NT) 
through the Programme de Bourses d'Excellence pour \'Etudiants \'Etrangers.
Computations were made 
on the supercomputer Guillimin, managed by Calcul Qu\'ebec and Compute Canada and funded by the Canada Foundation for Innovation (CFI), Minist\`ere de l'\'Economie, des Sciences et de l'Innovation du Qu\'ebec (MESI) and FRQ-NT. 
\bibliographystyle{JHEP}
\bibliography{biblio}

\end{document}